\documentclass[%
 reprint,
superscriptaddress,
 amsmath,
 amssymb,
 aps, 
]{revtex4-2}

\usepackage{graphicx}% Include figure files
\usepackage{dcolumn}% Align table columns on decimal point
\usepackage{bm}% bold math
\usepackage{xcolor} % colored text

\usepackage{enumitem}

\usepackage[colorlinks=true, allcolors=blue, breaklinks=true]{hyperref}

\begin{document}

\title{\textbf{Efficient measurement of neutral-atom qubits with matched filters} 
}% 

\author{Robert M. Kent}
\affiliation{The Ohio State University, Department of Physics, 191 West Woodruff Ave., Columbus, OH 43210, USA.}

\author{Linipun Phuttitarn}
\affiliation{Department of Physics, University of Wisconsin-Madison, 1150 University Avenue, Madison,
Wisconsin 53706, USA}

\author{Chaithanya Naik Mude}
\affiliation{Department of Computer Sciences, University of Wisconsin-Madison, 1210 W Dayton St, Madison, Wisconsin 53706, USA}

\author{Swamit Tannu}
\affiliation{Department of Computer Sciences, University of Wisconsin-Madison, 1210 W Dayton St, Madison, Wisconsin 53706, USA}

\author{Mark Saffman}
\affiliation{Department of Physics, University of Wisconsin-Madison, 1150 University Avenue, Madison,
Wisconsin 53706, USA}
\affiliation{Infleqtion, Inc., Madison, Wisconsin, 53703, USA}

\author{Gregory Lafyatis}
\affiliation{The Ohio State University, Department of Physics, 191 West Woodruff Ave., Columbus, OH 43210, USA.}

\author{Daniel J. Gauthier}
\email{Contact author: daniel.gauthier62@gmail.com}
\affiliation{ResCon Technologies, LLC, 1275 Kinnear Rd., Suite 239, Columbus, OH 43212, USA}

\noaffiliation

\date{\today}% It is always \today, today,

\begin{abstract}
Quantum computers require high-fidelity measurement of many qubits to achieve a quantum advantage. Traditional approaches suffer from readout crosstalk for a neutral-atom quantum processor with a tightly spaced array. Although classical machine learning algorithms based on convolutional neural networks can improve fidelity, they are computationally expensive, making it difficult to scale them to large qubit counts. We present two simpler and scalable machine learning algorithms that realize matched filters for the readout problem.  One is a local model that focuses on a single qubit, and the other uses information from neighboring qubits in the array to prevent crosstalk among the qubits. We demonstrate error reductions of up to 32\%  and 43\% for the site and array models, respectively, compared to a conventional Gaussian threshold approach. Additionally, our array model uses two orders of magnitude fewer trainable parameters and four orders of magnitude fewer multiplications and nonlinear function evaluations than a recent convolutional neural network approach, with only a minor (3.5\%) increase in error across different readout times. Another strength of our approach is its physical interpretability: the learned filter can be visualized to provide insights into experimental imperfections.  We also show that a convolutional neural network model for improved can be pruned to have 70$\times$ and 4000$\times$ fewer parameters, respectively, while maintaining similar errors. Our work shows that simple machine learning approaches can achieve high-fidelity qubit measurements while remaining scalable to systems with larger qubit counts.
\end{abstract}

\maketitle

\section{\label{sec:Introduction} Introduction}

Quantum computers promise to revolutionize fields such as cryptography \cite{Shor_1994,Bernstein_2017,Gidney_2021}, complex optimization \cite{Farhi_2014,Zhou_2020,optimization_nature_2024}, and drug discovery \cite{Cao_2018,Blunt_2022} due to their potential to solve important classes of problems faster than classical computers. However, we do not yet have a useful quantum computer because current machines have a small number of quantum bits (qubits) with errors that must be corrected.  These errors include imprecise quantum logic gate operations, qubit decoherence, and inaccurate qubit state measurement. Error correction schemes \cite{Shor_1995,Fowler_2012,Google_2024} are designed to detect and correct errors as they occur, but some of these protocols rely on fast qubit readout and feedback during computation. Therefore, there is a continuing need to develop fast and high-fidelity qubit measurement protocols.

An emerging platform for quantum computing uses an array of trapped neutral atoms \cite{Saffman_2010,Henriet_2020,Graham_2022,Manetsch_2024}, where a single laser-cooled atom is loaded into each far-off-resonance trap. The qubit states correspond most often to two hyperfine ground states of the atoms, which are manipulated by microwave or optical fields to realize gate operations.

The qubit states are measured using resonance fluorescence, where the atoms are illuminated with light that excites a cycling transition, causing the atoms in one of the hyperfine states to fluoresce. The fluorescence is then captured by a high-resolution, single-photon-sensitive digital camera, resulting in a grayscale image in which the atoms appear bright or dark, depending on their state. We denote the pixel-level image intensity by $I_{ij}$, where the index $ij$ denotes the pixel location.

There is great interest in determining whether classical machine learning algorithms can improve quantum computer technologies, especially the readout measurement for which the resulting information is in the classical domain. For example, Phuttitarn \textit{et al.} \cite{Phuttitarn_2024} recently demonstrated that a convolutional neural network (CNN) can process the classical intensities $I_{ij}$ to improve the accuracy of reading out a small (3 $\times$ 3) array of neutral atom qubits. They demonstrated a faster readout time to obtain the same measurement fidelity compared to traditional image-analysis methods for qubit readout.  A similar approach has been used for reducing errors in the readout of superconducting qubits \cite{Maurya_2023, Mude_2025}.

While CNNs are one of the leading machine learning algorithms for processing image data, the models tend to have millions of trainable parameters that must be optimized, requiring large training datasets.  In addition, once the model is learned, the complexity of the model requires substantial memory and computing resources to make a new prediction, making it difficult to deploy the model on a small computer chip embedded in the camera.

Here, we take an intermediate approach by applying a simpler machine learning algorithm to the same dataset described in Phuttitarn \textit{et al.} \cite{Phuttitarn_2024}.  The model has orders-of-magnitude fewer model parameters that are determined using fast linear optimization. Our model can be interpreted as a `matched filter,' where each $I_{ij}$ is multiplied by a weight $W_{ij}$ learned during training.  The weighted pixel intensities in a region around each trapped atom are summed and thresholded to estimate the qubit state.  The model weights are matched to the intensity probability distribution for each atom and compensate for system non-idealities to improve the state estimation. The fidelity of our algorithm is comparable to that of the more complex CNN algorithms and remains superior to traditional methods of processing readout images.

In the next section, we give background on the experiment, previous approaches for qubit readout algorithms, and the metrics for assessing model quality.  We then introduce our machine learning algorithm, present our results, and discuss the implications for our work.  

\section{Background} \label{Sec:Background}

\subsection{Experimental setup}

Our analysis is performed on the dataset measured and analyzed by Phuttitarn \textit{et al.} \cite{Phuttitarn_2024}.  Briefly, they load single cesium atoms in each of 9 far-off-resonance optical traps in a square 3$\times$3 array.  To read out the qubit state, all atoms in the array in one qubit state are pushed out of their trap using a large optical beam.  Then, resonant laser light is applied to the remaining atoms, which fluoresce.  Sites that are `dark' indicate atoms that were in the pushed-out state, and `bright' sites indicate atoms in the other state.

A high-numerical aperture lens images the fluorescence emitted by the trapped atoms onto an electron-multiplying charge-coupled device digital camera. They intentionally attenuate the light traveling along this path to simulate a noisy readout with a shorter exposure readout period and hence there is a reduced intensity for qubit readout.  Here, we focus on the setup where the array spacing is 5 $\mu$m, which is close enough that fluorescent light from one atom overlaps with its neighbors, causing crosstalk.  

% A high-numerical aperture lens images the fluorescence emitted by the trapped atoms onto an electron-multiplying charge-coupled device digital camera. They intentionally attenuate the light traveling along this path to simulate a setup where some of the light is used for other purposes and hence there is a reduced intensity for qubit readout.  Here, we focus on the setup where the array spacing is 5 $\mu$m, which is close enough that fluorescent light from one atom overlaps with its neighbors, causing crosstalk.  

They also measure the fluorescent light from the opposite side of the trap without purposeful attenuation, which gives a higher signal-to-noise ratio and hence a higher fidelity readout for a given measurement time.  This high signal-to-noise measurement path is used to determine the measurement fidelity for the first path and for training the machine-learning algorithms.  See Ref.~\cite{Phuttitarn_2024} for the details of the experimental setup.

\subsection{Traditional qubit readout}

The traditional approach for estimating the qubit state involves processing the pixel intensity data $I_{ij}$, which depends on the length of time the resonant optical fields illuminate the atoms.  Longer readout times achieve higher state fidelity (defined below), but this slows down possible error correction protocols and the experimental cycle time.

The simplest approach is to find
\begin{equation}
S_{\rm square} = \sum_{ij}^N I_{ij},
\label{eq:square_filter}
\end{equation}
where the indices are taken from a square surrounding each atom with $N$ pixels.  The central atom location can be determined by averaging many images of the array when the atoms are in different qubit states.  The probability density of the sum $\mathcal{P}(S_{\rm square})$ for the dark and bright states are each expected to be approximately Gaussian, and the optimum threshold for the state estimate is the value of $S_{\rm square}$ where the distributions intersect.  For short readout times, the signal-to-noise of the images are low and there is substantial overlap in the distributions, which causes state misidentification.

A lower error algorithm assumes that the intensity profile from each atom has a Gaussian shape so that the sum takes the form
\begin{equation}
S_{\rm Gaussian} = \sum_{ij}^N W_{ij} I_{ij}.
\label{eq:gaussian_filter}
\end{equation}
Here, $W_{ij}$ are discrete values of a two-dimensional Gaussian function centered at the location of each atom, where the standard deviation $\sigma$ is determined by fitting a circular Gaussian to the averaged images. In Ref. \cite{Phuttitarn_2024}, these weights are defined across the entire image but remain small far from the center, where they can often be set to zero to reduce computational cost.

The key advantage of these approaches is that they do not require knowing the true qubit states.  These algorithms are called `unsupervised' in the machine learning field because they do not need `ground truth' data for training (optimizing) the algorithm.  However, because the filters are local to each qubit, they struggle to account for crosstalk.

\subsection{Readout measurement metrics}

One key metric to quantify the qubit state readout performance is the classification fidelity given by
\begin{equation}
    \mathcal{F} = 1 - \frac{P(B_p | D) + P(D_p | B)}{2},
\label{eq:fidelity}
\end{equation}
where $P(B_p | D)$ is the probability that the model predicts the qubit to be in a bright state given that it was actually dark (false-bright), and $P(D_p | B)$ is the probability that the model predicted the qubit to be in a dark state given that it was actually bright (false dark).

Next, we define the cross-fidelity \cite{Google_2021}, which measures the predicted correlations between neighboring qubits, giving insight into the model's susceptibility to crosstalk. The cross-fidelity is given by 
\begin{equation}
    \mathcal{F}_{kl}^{\text{CF}} = 1 - \langle P(D_k | B_l) + P(B_k | D_l)\rangle,
\label{eq:cross_fidelity}
\end{equation}
where $P(D_k | B_l)$ is the probability that the model predicted the qubit at site $k$ as dark given that the model predicted bright at site $l$, and $P(B_k | D_l)$ is the probability that the model predicted the qubit at site $k$ as bright given that the model predicted dark at site $l$. 

Lastly, we define the infidelity reduction, which represents the percent decrease in incorrect state predictions relative to the traditional Gaussian filter approach, as 
\begin{equation}
\eta = \frac{(1 - \mathcal{F_\sigma}) - (1 - \mathcal{F}_{\text{MF}})}{1 - \mathcal{F_\sigma}}
\label{eq:infidelity_reduction}
\end{equation}
where $\mathcal{F_\sigma}$ is the fidelity given by the traditional Gaussian filter approach, and $\mathcal{F}_{\text{MF}}$ is the fidelity given by the matched filter approach discussed in the next section.

\section{Enhancing qubit readout using classical machine learning}

Recently, Phuttitarn \textit{et al.} \cite{Phuttitarn_2024} demonstrated that a CNN can mitigate crosstalk and increase measurement fidelity. This is achieved by learning nonlinear filters (their so-called CNN-Site model) or by processing a larger pixel area to account for neighboring qubits (CNN-Array model). 

A CNN is a feed-forward artificial neural network, where information flows in a unidirectional manner from input to output \cite{Goodfellow_2016}.  It has many layers, each containing many `neurons,' where a neuron performs a nonlinear transformation of weighted and summed input information.  Common nonlinear transforms include sigmoids or rectifying piecewise-linear functions.  Small filters are applied to the input image data to form a convolution between the image and the filter; using several filters focuses model attention on different image features.  The output of the network is mapped to numerical `classes' representing the problem solution: identifying the dark (mapped to 0) and bright (mapped to 1) qubit states.

This approach typically relies on supervised learning, which requires that the true qubit labels (dark or bright) are known to train the model. In this work and Ref. \cite{Phuttitarn_2024}, the true labels are determined by applying a Gaussian filter to the unattenuated path, but the machine learning model is trained to predict qubit states using only images from the attenuated path.

Each dataset that corresponds to a different measurement time is split into training, validation, and testing images: the training images are used for training or optimizing the neural network weights, the validation images are used to monitor the performance during training, and the testing set is used to evaluate the model's performance on unseen data.

Nonlinear optimization algorithms are required for training the weights because they appear within nonlinear functions. Nonlinear optimization is computer intensive, often requiring hours of computing time using graphical processor units.  Furthermore, the models tend to have many weights: their CNN-Array model requires tens of millions of parameters and 30 passes through the dataset during optimization. 

% should to mention matched filters here (is matched filter defined earlier?)
Here, we develop two supervised machine learning algorithms that realize matched filters.  Our approach dramatically reduces the computational complexity of the model, both in the training and evaluation phases.  They are linear models similar to the Gaussian filter (Eq. \ref{eq:gaussian_filter}), but the weights $W_{ij}$ are learned using linear least-square regression. This machine learning model is faster to train because it only requires linear convex optimization with a guaranteed solution. 

\begin{figure}[h!]
\includegraphics[width=\columnwidth]{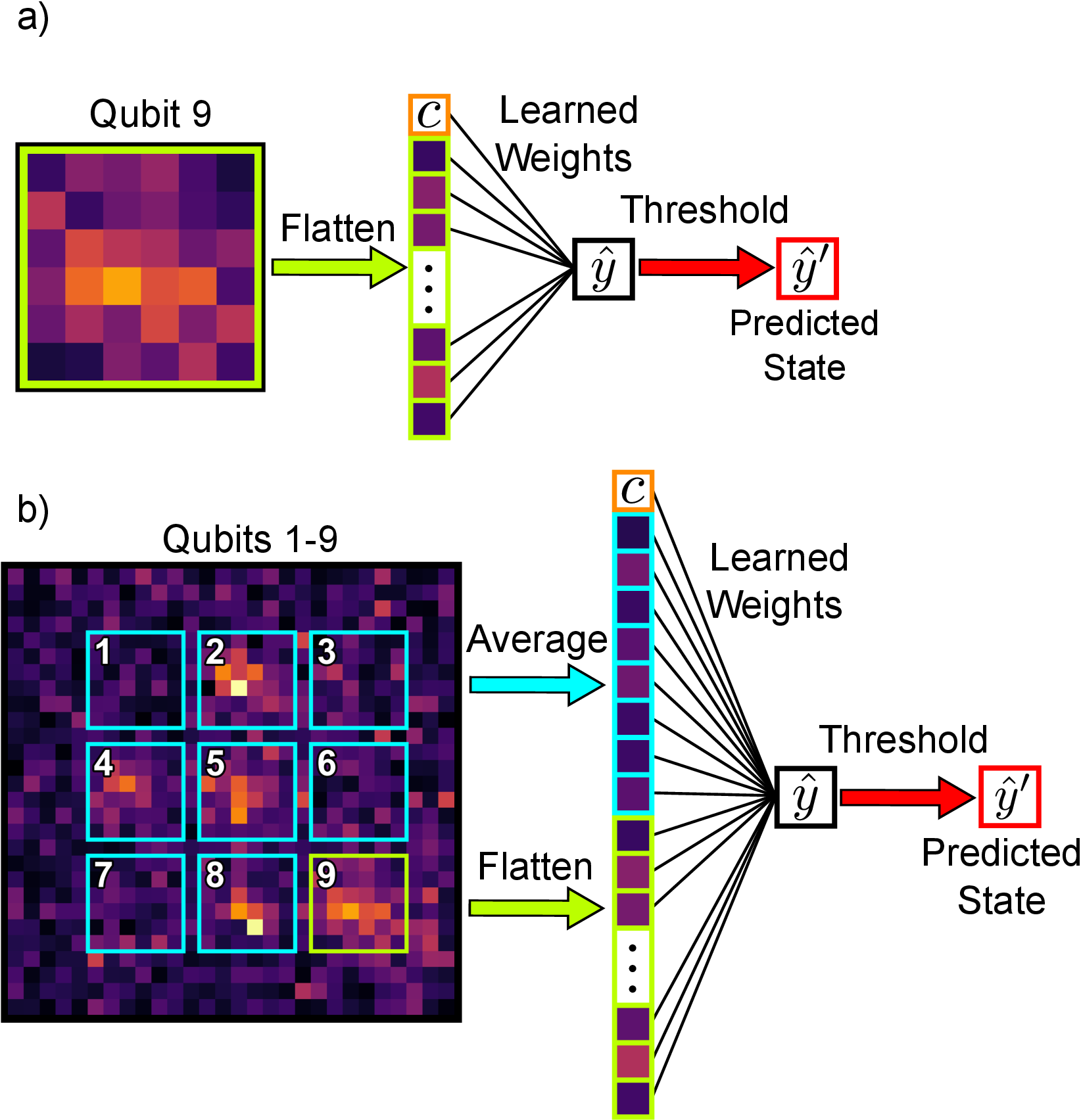}
\caption{\label{fig:MF_diagrams} Matched filter designs for a single qubit with boundary size $s=6$. (a) MF-Site configuration. (b) MF-Array configuration.}
\end{figure}

\subsection{\label{sec:MF-Site} MF-Site}

% need to be clear that this predicts the state of one qubit, not all at once.
The MF-Site model classifies the state of a single qubit using only its corresponding pixel data, as illustrated in Fig. \ref{fig:MF_diagrams}a for an example image where qubit 9 is in a bright state. The pixel data for this qubit is determined by first finding the center of each qubit in the image (see Appendix \ref{sec:Preprocessing}), and then placing a square boundary with width $s$. The pixel intensities $I_{ij}$ inside each boundary are augmented by a constant $c$ and placed in a flattened data structure known as the feature vector, depicted in Fig. \ref{fig:MF_diagrams}a. 

A dot product is formed by the feature vector and a vector containing the weights $W_{ij}$, which gives the predicted state $\hat{y}$.  Here, $\hat{y}$ is continuous but should be close to 0 for a dark state and 1 for a bright state. Last, a threshold is applied to $\hat{y}$ to produce a binary classification $\hat{y}' \in [0,1]$, corresponding to the state estimate. Details about the numerical procedure for determining the weights and optimal threshold are given in Appendix \ref{sec:Training}. 

\subsection{\label{sec:MF-Array} MF-Array}

The MF-Array model also classifies the state of a single qubit, but it also uses information from neighboring qubits to detect crosstalk, as depicted in Fig. \ref{fig:MF_diagrams}b.  In this example image, the goal is to classify the state of qubit 9 when qubits 1, 3, 6, and 7 are in the dark state, and qubits 2, 4, 5, 8, and 9 are in the bright state. 

Boundaries are placed around each qubit and the pixels inside the boundary corresponding to the target qubit are flattened to form part of the feature vector, as in the CNN-Site model. Pixels inside the boundaries around the other qubits are averaged and included in the feature vector, providing the model with information about the neighbor states. Since averaging primarily involves addition and introduces only a few additional values to the feature vector, the MF-Array has only a slight increase in computational complexity compared to the MF-Site model for a given boundary size. The weights and threshold are optimized similarly to the MF-Site model, with details given in Appendix \ref{sec:Training}.

Including the average pixel values from neighboring qubits allows the model to detect crosstalk, as a high average from a neighboring qubit suggests it is in the bright state, contributing more fluorescence that can reach the target qubit’s pixels. By accounting for this effect, the model can better distinguish the target qubit’s true state from crosstalk interference.

The advantages of this approach are that it can detect crosstalk with minimal computational overhead, and significantly reduces the amount of image data the model must process. Unlike other array-based methods that analyze entire pixel grids, this model condenses relevant information into just a few averaged values, capturing neighboring qubit influences without directly processing all surrounding pixels.

\section{\label{sec:Results} Results}

In this section, we compare the classification performance, crosstalk mitigation, and discuss the physical insights provided by the learned weights of the matched filter. We also compare the computational complexity of the MF-Site and MF-Array to the traditional approaches and Phuttitarn's CNN algorithms \cite{Phuttitarn_2024}.

\subsection{\label{sec:Classification performance} Classification performance}

\begin{figure}[h]
\includegraphics[width=\columnwidth]{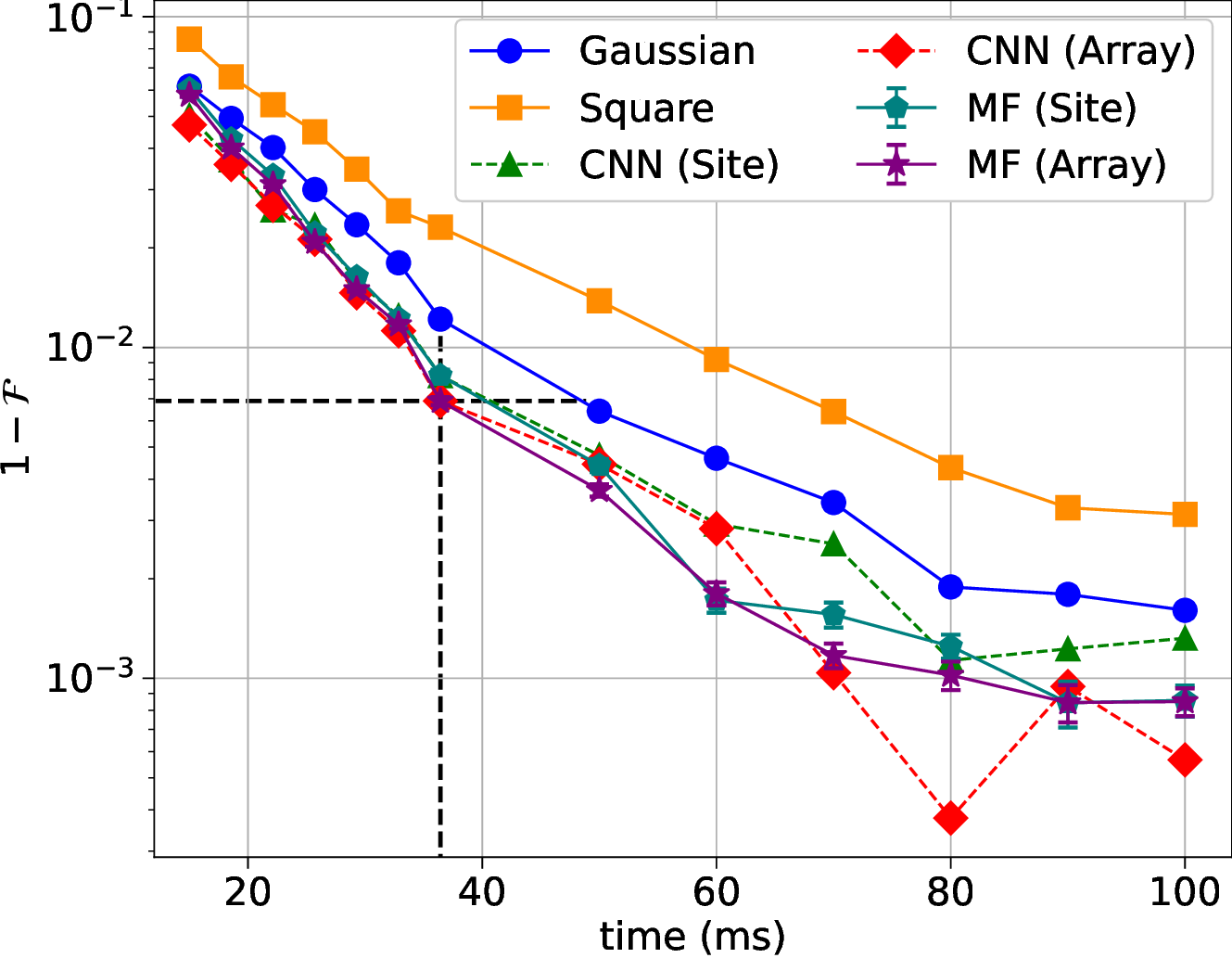}
\caption{\label{fig:infidelity_vs_time} The measurement infidelity from different analysis methods at different readout times. The square, Gaussian, CNN (site) and CNN (array) infidelity values are taken from Ref. \cite{Phuttitarn_2024}.}
\end{figure}

\begin{figure}[h]
\includegraphics[width=\columnwidth]{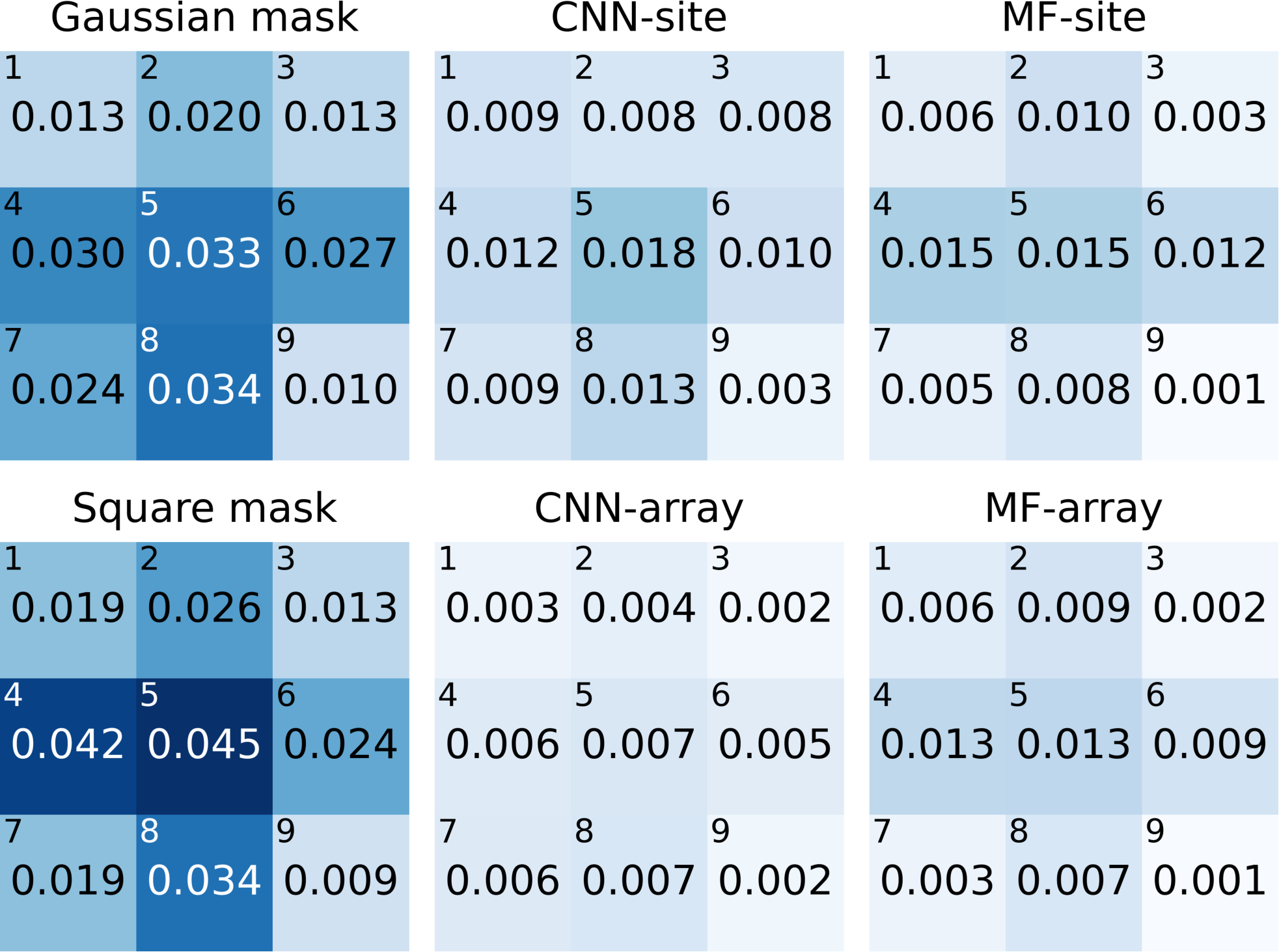}
\caption{\label{fig:infidelity_grid_36ms} Classification infidelity for each qubit (numbered in top left of each box) for a measurement time of 36 ms. The standard error for the matched filter values is below 0.0001 for each site.}
\end{figure}

\begin{table*}[!htbp]
\caption{\label{tab:table_1} Comparison of cross-fidelity values at a measurement time of 36 ms for different methods. The mean standard errors for the MF-Site and MF-Array are 0.0004 for the C-NN values and 0.0002 for the E-E values. The boundary size $s$ is optimized for both the MF-Site and MF-Array.}
\begin{ruledtabular}
\begin{tabular}{llcccccc}
\textbf{C-NN} & & \textbf{Gaussian} \cite{Phuttitarn_2024} & \textbf{Square} \cite{Phuttitarn_2024} & \textbf{CNN-site} \cite{Phuttitarn_2024} & \textbf{CNN-array} \cite{Phuttitarn_2024} & \textbf{MF-Site} & \textbf{MF-Array} \\
\hline
1 & $\mathcal{F}^{\text{CF}}_{52}$ & 0.0298 & 0.0524 & 0.0151 & 0.0099 & 0.0108 & 0.0102 \\
2 & $\mathcal{F}^{\text{CF}}_{54}$ & 0.0211 & 0.0298 & 0.0146 & 0.0099 & 0.0058 & 0.0066 \\
3 & $\mathcal{F}^{\text{CF}}_{56}$ & 0.0218 & 0.0284 & 0.0120 & 0.0022 & 0.0071 & 0.0074 \\
4 & $\mathcal{F}^{\text{CF}}_{58}$ & 0.0392 & 0.0607 & 0.0120 & 0.0020 & 0.0096 & 0.0077 \\
5 & $\langle |\mathcal{F}^{\text{CF}}_{5l}| \rangle$ & \textbf{0.0280} & \textbf{0.0428} & \textbf{0.0134} & \textbf{0.0060} & \textbf{0.0083} & \textbf{ 0.0080} \\
\hline
\textbf{E-E} & & \textbf{Gaussian} \cite{Phuttitarn_2024} & \textbf{Square} \cite{Phuttitarn_2024} & \textbf{CNN-site} \cite{Phuttitarn_2024} & \textbf{CNN-array} \cite{Phuttitarn_2024} & \textbf{MF-Site} & \textbf{MF-Array} \\
\hline
6  & $\mathcal{F}^{\text{CF}}_{13}$ &  0.0110  & 0.0139  & 0.0205  & 0.0110  & 0.0104 & 0.0090 \\
7  & $\mathcal{F}^{\text{CF}}_{79}$ & -0.0036  & -0.0005 & -0.0094 & -0.0036 & -0.0063 & -0.0090 \\
8  & $\mathcal{F}^{\text{CF}}_{17}$ &  0.0038  & 0.0091  & -0.0016 &  0.0038 & -0.0006 & -0.0027 \\
9  & $\mathcal{F}^{\text{CF}}_{39}$ &  0.0028  & 0.0090  &  0.0031 &  0.0028 & 0.0002 & -0.0009 \\
10 & $\mathcal{F}^{\text{CF}}_{19}$ & -0.0068  & -0.0043 & -0.0098 & -0.0068 & -0.0077 & -0.0089 \\
11 & $\mathcal{F}^{\text{CF}}_{37}$ &  0.0070  & 0.0078  &  0.0080 &  0.0070 & 0.0063 & 0.0056 \\
12 & $\langle |\mathcal{F}^{\text{CF}}_{kl}| \rangle$ & \textbf{0.0058} & \textbf{0.0074} & \textbf{0.0087} & \textbf{0.0058} & \textbf{0.0052} & \textbf{0.0060} \\
\end{tabular}
\end{ruledtabular}
\end{table*}

\begin{table*}[!htbp]
\caption{\label{tab:table_2} Comparison of the number of trainable parameters, evaluation complexity, and fidelity of different approaches. The standard error of the mean fidelity is 0.0003 for both the MF-Site and MF-Array.}
\begin{ruledtabular}
\resizebox{\textwidth}{!}{%
\begin{tabular}{lccccc}
\textrm{Model} & \textrm{\# Trainable Param.} & \textrm{\# Mult.} & \textrm{\# Nonlinear Functions} & \textrm{Mean Fidelity} \\
\hline
Square       & 0   & 0       & 0      & 0.9712 \\
Gaussian     & 18   & 855 ($2.5 \times 10^5$)  & 0      & 0.9804 \\
CNN-Site     & $3.6 \times 10^5$   & $1.9 \times 10^7$       & $5.9 \times 10^4$  & 0.9856 \\
CNN-Array    & $7.5 \times 10^7$   & $1.2 \times 10^8$       & $1.2 \times 10^5$   & 0.9866 \\
\textbf{MF-Site ($s=opt.$)} & \textbf{909-1731} & \textbf{909-1731} & \textbf{0} & \textbf{0.9841}  \\
\textbf{MF-Array ($s=opt.$)} & \textbf{522-1576} & \textbf{522-1576} & \textbf{0} & \textbf{0.9851}  \\
\end{tabular}
} % End of resizebox
\end{ruledtabular}
\end{table*}

Figure \ref{fig:infidelity_vs_time} shows the infidelity as a function of the qubit measurement time.  The general trend is that infidelity decreases for longer measurement times as mentioned above.  The goal is to reduce the measurement time for a given infidelity threshold. 

For the traditional filters, we see that the square filter performs the worst (larger infidelity for any measurement time), whereas the Gaussian filter has a marked improvement because it accounts for the expected intensity profile of the fluorescent light emitted by each qubit.

We see that all four classical machine learning algorithms substantially improve upon the traditional Gaussian filter algorithm.  These algorithms perform equally well to within the apparent variation in the data, although the matched filter occasionally outperforms the CNN algorithms.

The error bars for the matched filter approaches represent the standard error, calculated from infidelity variations across 10 different random shuffles of the dataset. This ensures more representative fidelity curves, as single misclassifications at longer readout times, such as the CNN-Array at 80 ms, can cause large fidelity jumps that do not reflect overall performance. See Section \ref{sec:uncertainty} for details.

The vertical black dashed line indicates a measurement time of 36ms, at which Phuttitarn \textit{et al.} \cite{Phuttitarn_2024} report maximum infidelity reductions for their CNN-site and CNN-array of 32\% and 43\%, respectively, relative to the Gaussian approach. At this measurement time, the MF-site and MF-array approaches have matching infidelity reductions of 32\% and 43\%, respectively.

The horizontal black dashed line indicates the point where Phuttitarn et al. \cite{Phuttitarn_2024} report the largest reduction in measurement time, with CNN-site and CNN-array achieving 20\% and 25\% reductions, respectively, relative to the Gaussian approach. At the same fidelity threshold, the MF-site and MF-array methods have matching readout time reductions of 20\% and 25\%, respectively.

We also see that the MF-Array model generally matches or outperforms the MF-Site approach, especially for longer readout times where crosstalk is more likely. This indicates that including the averages of neighboring qubit pixel intensities in the feature vector can improve performance.

Figure \ref{fig:infidelity_grid_36ms} shows the classification infidelity for each qubit using each approach at a measurement time of 36 ms. As before, the goal is to minimize the infidelity for each qubit, but the center qubit (site 5) is particularly important since it interacts with the most neighbors, serving as an indicator of how performance scales in larger qubit systems.

We see that both the Square and Gaussian approaches have significantly higher infidelity across all qubits compared to the other approaches, with the most significant increases being for the central qubit and its nearest neighbors. This highlights that the traditional approaches cannot account for crosstalk.

Looking at the other approaches, we see that CNN-Site, MF-Site, and MF-Array perform similarly well, all improving center qubit performance over traditional methods, with MF-Array achieving the lowest infidelity of the three. The CNN-Array further reduces the infidelity for the center qubit and its nearest neighbors but comes with a significant increase in computational complexity compared to the CNN-Site, as discussed later.

Table \ref{tab:table_1} shows the cross-fidelity (Eq. \ref{eq:cross_fidelity}) for the center qubit to its nearest neighbors (C-NN) and for the edge sites (E-E) at a measurement time of 36 ms. Positive values for a given qubit pair indicate that the predictions of their states are correlated, while a negative value signify anti-correlation. Ideally, the cross-fidelity magnitude should be close to zero for a given qubit pair, indicating that the model is mitigating the effects of crosstalk. 

We see that for the nearest neighbors, both the MF-Site and MF-Array approaches achieve average cross-fidelity magnitudes over three times lower than those of the Gaussian approach, and lower than the CNN-Site approach. While the CNN-Array has the lowest cross-fidelity for the nearest neighbors, the MF-Array is not significantly higher and shows an improvement over the MF-Site.

Looking at the edge sites, we see that the cross-fidelity magnitude is typically much smaller compared to the magnitude of the nearest neighbors for most approaches. This aligns with expectations as edge sites are spaced further apart, reducing the likelihood of crosstalk. While cross-fidelity varies between approaches, the overall lower crosstalk at edge sites makes these differences negligible.

Given these cross-fidelity metrics, we conclude that for nearest neighbors, both the MF-Site and MF-Array significantly reduce crosstalk compared to the standard approaches while achieving performance comparable to the CNN methods. Notably, the additional features in MF-Array that incorporate information about neighboring qubit states further reduce crosstalk compared to MF-Site.

\subsection{\label{sec:Physical interpretability of weights} Physical interpretability of weights}

\begin{figure}[h]
\includegraphics[width=\columnwidth]{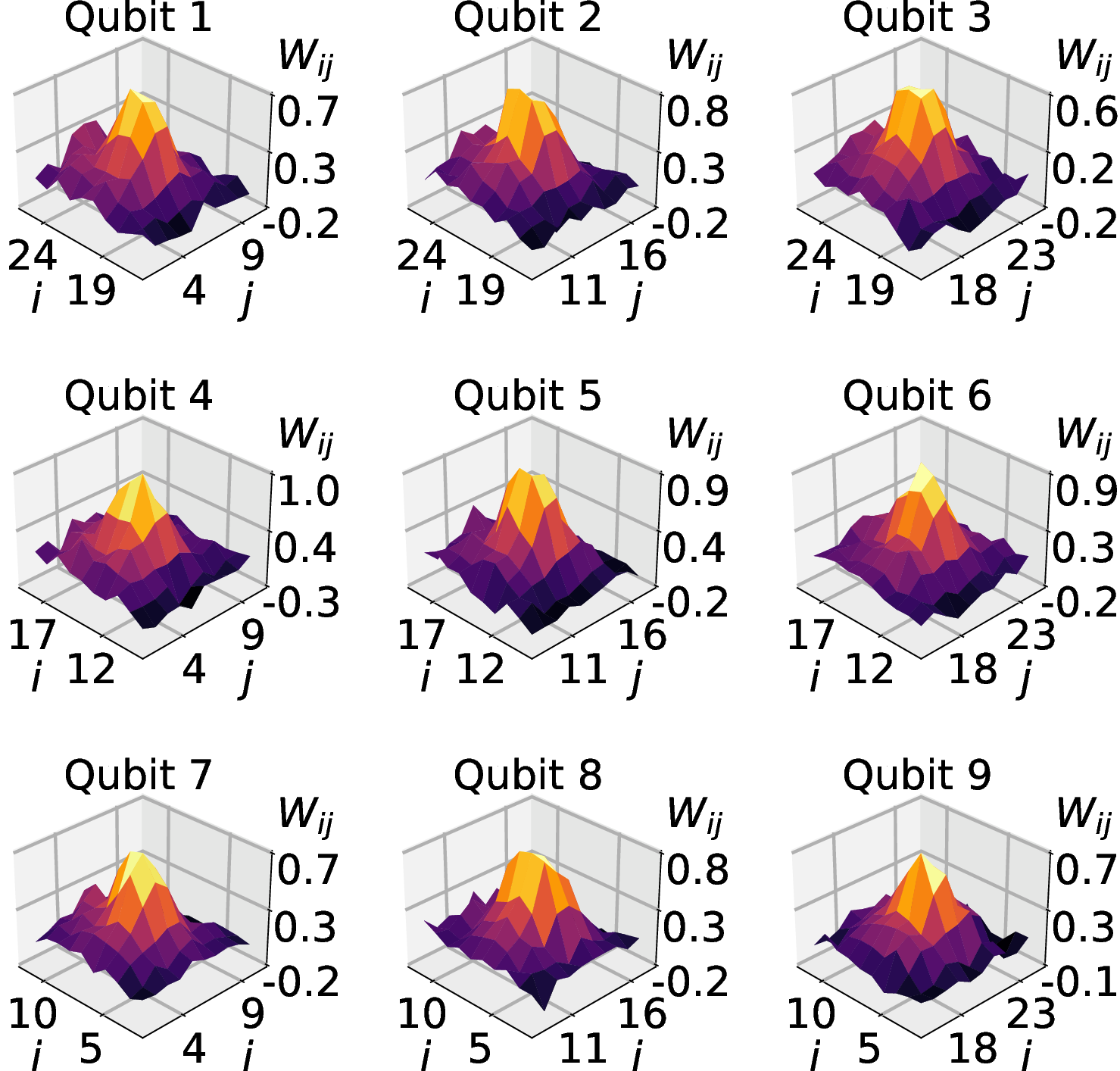}
\caption{\label{fig:false_3D_plots} False-3D images of the learned weights for each qubit.}
\end{figure}

One strength of the matched filter approach is that the learned weights provide physical insights into the optimal filter. Figure \ref{fig:false_3D_plots} shows the weights for each qubit in the MF-Site model with a boundary size of $s=10$ at a measurement time of 36 ms. The $i$ and $j$ axes represent the pixel coordinates, and the vertical scale represents the weight values $W_{ij}$. At this boundary size, qubit regions overlap, allowing the model to account for crosstalk to some degree.

We see that the weights have a Gaussian-like shape, but the weights become increasingly bumpy as the distance from the center increases for some qubits. We hypothesize that the optimal filter weights deviate from a Gaussian to account for imperfections in the experimental setup, such as optical aberrations or pixel quantum efficiency variation. 

The variation in the weights near the boundary of each region is affected by the least-squares regularization parameter $\alpha$.  Optimizing $\alpha$ has only a minor effect on model prediction accuracy, so we set it equal to zero (normal least-squares optimization) for simplicity. Increasing $\alpha$ imposes a penalty on large weights, which smooths out the bumps and produces a more Gaussian-like shape.

\begin{figure}[h]
\includegraphics[width=\columnwidth]{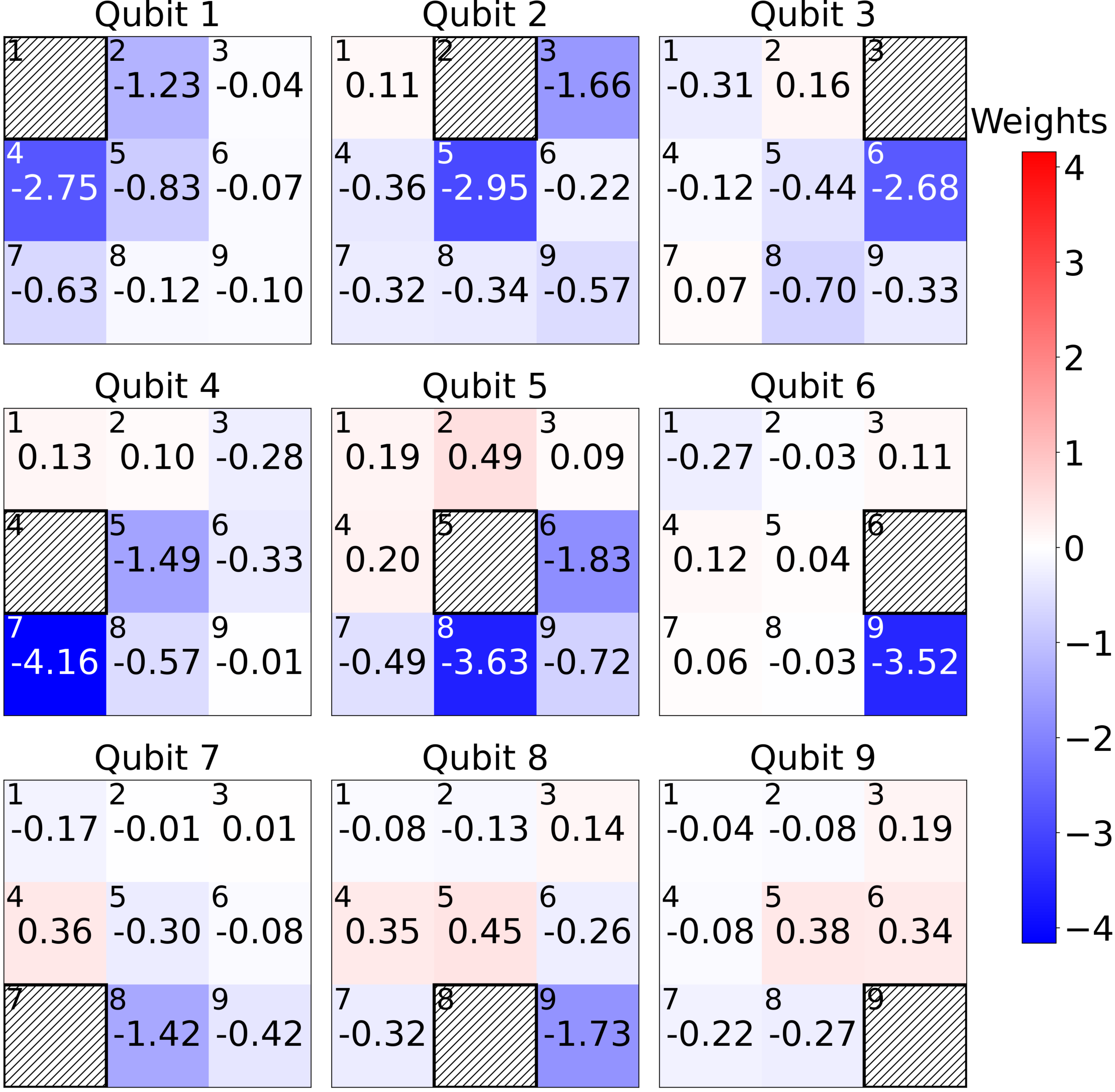}
\caption{\label{fig:neighbor_weights} Weights of neighbor qubits (numbered in the top left of each box) for the MF-Array model.}
\end{figure}

We can also visualize the weights that correspond to the neighboring qubits in the MF-Array approach, shown in Figure \ref{fig:neighbor_weights} for the optimal boundary size $s=8$ at a readout time of 36 ms. The general trend is that the nearest neighbor qubits, particularly those below and to the right of each qubit, exhibit large negative weight values, indicating that the model is accounting for crosstalk from those qubits. In contrast, the qubits above and to the left of the target qubit often show small, near-zero positive weights. This disparity may reflect imperfections in the optical setup, causing the fluorescence from neighboring qubits to propagate in specific directions. The additional physical interpretability provided by the MF-Array may help to identify and address such setup-related imperfections.

\subsection{Algorithm computational complexity}

Next, we compare the computational complexity of the approaches by considering the number of trainable parameters for each model and the number of multiplications and nonlinear functions required during evaluation. Table \ref{tab:table_2} summarizes these factors and also includes the mean classification fidelity across all qubits and measurement times for reference.

We see that the traditional square filter approach requires no trainable parameters, multiplications, or nonlinear functions, making it the simplest to evaluate but it results in the lowest mean fidelity. The Gaussian approach has 18 trainable parameters which correspond to the standard deviation and amplitude of the Gaussian fit for each qubit. As mentioned before, in Ref. \cite{Phuttitarn_2024} the Gaussian filter for each qubit spans the entirety of the raw $442 \times 62$-pixel image, bringing the total number of multiplications to $2.5 \times 10^5$. However, this is not practical for real-time evaluation. Instead, we estimate the number of multiplications for the Gaussian filter by selecting only the values in the filter greater than $10^{-3}$ within its 0 to 1 range. Despite their lower performance, these traditional approaches scale linearly with the number of qubits, making them practical for systems with a large number of qubits.

% Mark had a comment: "A single CNN site and an MF site are used to classify all qubits in the array. Therefore, neither method should scale linearly with the number of qubits." But I think the CNN site only classifies one qubit in the array. Also, the MF-Site only classifies one qubit at a time
In contrast, the CNN-site and CNN-array approaches require many orders of magnitude more trainable parameters, multiplications, and nonlinear functions compared to the traditional approaches, but offer significant increases in fidelity. While the CNN-site model scales linearly with the number of qubits, evaluating it for even a single qubit can exceed the resources of a single chip. In its current form, the CNN-array scales non-linearly with the number of qubits because the input image size must increase to include the additional qubits. This may not require adjustments to the number or shape of the convolutional layers, but it would greatly increase the number of neurons in the final layers of the model. The CNN-array model could be modified to predict only the center qubit state while keeping the same input pixel size, but this would still require a substantial computational load during both training and evaluation. 

The complexity metrics for the matched filter approaches are presented as ranges, as the boundary size $s$ is optimized for each qubit and measurement time (see Appendix \ref{sec:Validation}). In some cases, the MF-Array uses fewer trainable parameters and multiplications than the MF-Site, despite the MF-Array's ability to process multiple regions of the image. This is likely because the optimal MF-Site tends to require larger boundary sizes $s$ to capture information from neighboring qubits, whereas the MF-Array inherently incorporates this information by design.

The matched filters require a larger number of parameters than the traditional approaches, but two (four) orders of magnitude fewer than the CNN-Site (CNN-Array) approaches. Moreover, the linear optimization used to determine these parameters is significantly faster and requires fewer computational resources.

The number of multiplications required for the matched filter approaches is slightly larger on average than the Gaussian filter, but this is only because the optimal boundaries include more pixels. In some cases, the optimal MF-Array uses less multiplications than the Gaussian. Like the Gaussian approach, the matched filters do not require the evaluation of any nonlinear functions. Despite the matched filters requiring four (five) orders of magnitude fewer multiplications and nonlinear function evaluations than the CNN-Site (CNN-Array) approaches, they achieve similar mean fidelities.

Critically, the number of operations in the MF-Site approach scales linearly with the number of qubits, and in the MF-Array approach, limiting the feature vector to include only nearest neighbors also results in linear scaling. In addition, averages can be converted to sums, with divisions absorbed into the weights, further reducing the number of operations.

In Appendix \ref{sec:Pruning}, we prune the CNN-Site and CNN-Array approaches to simplify these models, reducing the number of parameters by 70$\times$ and 4000$\times$, respectively. The pruned CNN-Array uses single-site CNNs as inputs, significantly improving the complexity scaling with increasing qubit number. However, the pruned CNNs are still significantly more complex than the matched filters and provide similar prediction accuracies.

\section{\label{sec:Conclusions} Discussion}

Our algorithms require a set of multiplications that scale linearly with the number of qubits and hence can be applied to large qubit arrays.  Furthermore, the multiplications can be done in parallel using, for example, a field-programmable gate array, where the computations can be executed in a few clock cycles.  This will allow the qubit state readout computations to be completed on the nanosecond time scale, which is negligible compared to the measurement time.  

Learning the matched filter requires only regularized least-square regression, which can be done one feature vector at a time using sequential regression methods \cite{Hertz_1991}.  
This will allow for on-chip learning without storing the dataset or transferring it to another computer for processing. Constantly updating the weights will allow the matched filter to follow changes in the experimental setup.

\begin{acknowledgments}
We gratefully acknowledge the financial support of the U.S. Air Force Office of Scientific Research, Contract \#FA9550-22-1-0203. The work at UWM was supported by 
the U.S. Department of Energy Office of Science National Quantum Information Science Research Centers as part of the Q-NEXT center,  and NSF award No. 2210437.
\end{acknowledgments}

\appendix

\section{\label{sec:Methods} Methods}

\subsection{\label{sec:Preprocessing} Preprocessing}

The first step is to crop the full 442$\times$62-pixel image to obtain the 28$\times$28-pixel image corresponding to the secondary path. For each measurement time, there are 6,002 images, randomly shuffled and split into 60\% for training, 20\% for testing, and 20\% for validation. Next, we find the mean pixel intensity across all training images and subtract it from all images.  Finally, we determine the maximum range of intensities across all images and normalize the data by this value. 

For all algorithms, we must determine the location of each qubit in the image. This is accomplished by averaging all images in the training dataset, which produces an image with bright spots in the location of each qubit. We use the \texttt{peak\_local\_max} function from the \texttt{skimage.feature} package to determine an estimate of the bright spots in the image, and then use these coordinates as initial guesses for Gaussian fits using \texttt{curve\_fit} from the \texttt{scipy.optimize} package. The fitted coordinates are used as the center for the boundaries for each qubit. 

\subsection{\label{sec:uncertainty} Estimating Statistical Uncertainty in Infidelity}

As mentioned in the previous section, the images are split and randomly shuffled into training, testing, and validation sets. For Figures \ref{fig:infidelity_vs_time} and \ref{fig:infidelity_grid_36ms}, and Tables \ref{tab:table_1} and \ref{tab:table_2}, the dataset is split using 10 independent random shuffles, the infidelities at each measurement time are computed, and the infidelities for each random shuffle are averaged. The error bars and quoted uncertainties represent the standard error, or the standard deviation of the values over the 10 random shuffles, and then divided by $\sqrt{10}$. These uncertainties do not account for any systematic uncertainties present in the experiment. 

\subsection{\label{sec:Training} Training}

In the training step, the feature vectors for each image $\mathbf{x} \in \mathbb{R}^{d \times 1}$ (see Fig. \ref{fig:MF_diagrams}), where $d$ is the number of elements in the feature vector, are placed into a matrix $\mathbf{X} \in \mathbb{R}^{d \times M}$, where $M$ is the number of images in the training dataset, and the corresponding ground truth labels from the primary path $y \in [0,1]$ are placed into a matrix $\mathbf{Y} \in \mathbb{R}^{1 \times M}$. 

The weights $\hat{\mathbf{W}}$ are learned using linear regression with an L2 regularization, also known as Tikhonov regularization \cite{Vogel_2002}, which has a guaranteed solution given by
\begin{equation}
    \hat{\mathbf{W}} = \mathbf{Y} \mathbf{X}^\top \left( \mathbf{X} \mathbf{X}^\top + \alpha \mathbf{I} \right)^{-1}
\label{eq:ridge_regression}
\end{equation}
where $\alpha$ is the regularization parameter.  Regularization penalizes the size of the weights and compensates for possible ill-conditioning of the matrix $\mathbf{X}\mathbf{X}^T$. 

We evaluate Eq. \ref{eq:ridge_regression} for each qubit and measurement time, setting $\alpha=0$, and store the corresponding weights. We also compute the weights for different boundary sizes $s$ ranging from 2 to 14 in steps of 1. The optimization the threshold and the boundary size are detailed in the next section.

\subsection{\label{sec:Validation} Optimizing model metaparameters}

For each qubit and measurement time, we simultaneously optimize the threshold and boundary size. For each boundary size, the weights learned in the training step are used to predict the qubit states for all feature vectors $\mathbf{x}$ in the validation set by computing $\hat{y}=\hat{\mathbf{W}} \mathbf{x}$. Next, $\hat{y}'$ is computed for all thresholds ranging from 0.01 to 0.99 in steps of 0.01. Lastly, the combination of boundary size and threshold are chosen that maximize the classification fidelity on the validation set.

\begin{figure*}[htbp]
\centering
\includegraphics[width=\textwidth]{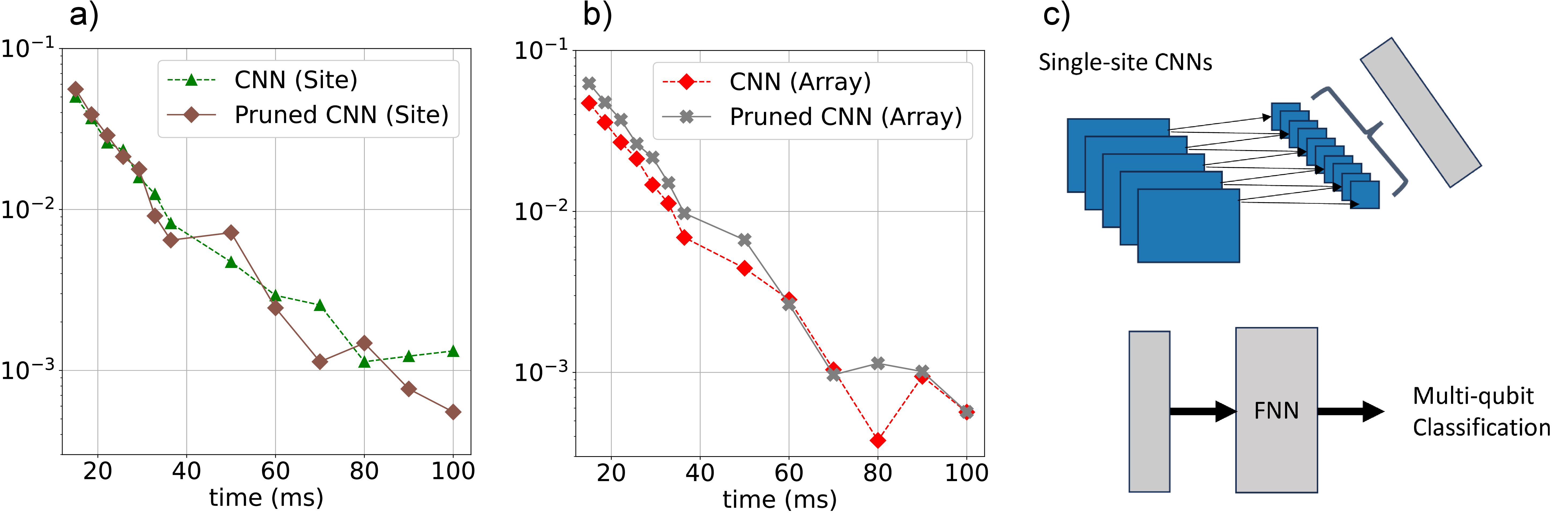}
\caption{(a) Infidelity comparison between CNN-Site (Baseline-355k Model) and Pruned CNN-Site (5k Model) (b) Infidelity comparison between CNN-Array (Baseline-74M Model) and Pruned CNN-Array (18k Model) (c) Overview design of hierarchical models using $N$ single-site CNN models for $N$ qubits and using additional FNN as higher-level layer.}
\label{fig:pruned_infidelity}
\end{figure*}

\subsection{\label{sec:Testing} Testing}

Lastly, the optimal boundary size, weights, and thresholds for each qubit are used to obtain $\hat{y}'$ for each image in the testing image set and compared to the ground truth values $y$ to obtain the classification fidelity. 

\subsection{\label{sec:Cross-Fidelity Calculation} Cross-Fidelity Calculation}

To obtain a more representative cross-fidelity calculation, we apply the trained models to a larger dataset of 39,976 images instead of the typical 1,200-image test sets. This is because the expanded dataset includes more examples of the possible permutations of the 9-qubit states, allowing for a more comprehensive assessment of crosstalk effects and improving the reliability of the reported cross-fidelity metrics.

\section{\label{sec:Pruning} Efficient CNN Architectures for Real-Time Qubit Readout}

This Appendix addresses whether a reduced-complexity CNN can approach the size of the matched filter described in the main text while maintaining the prediction accuracy. Our conclusion is that the models can be substantially reduced with a minor performance penalty, but they are still much larger than the matched filter models. It is important to document our efforts to reduce the complexity of the CNN models because they represent the current state-of-the-art in the ML field.

Standard approaches for reducing CNN model size include pruning, quantization, and architecture simplification. Pruning eliminates redundant weights or filters, quantization reduces numerical precision (\textit{e.g.}, from 32-bit float to 8-bit integer), and architecture simplification uses efficient structures such as depthwise separable convolutions (\textit{e.g.}, MobileNet) to minimize computation with minimal accuracy degradation. These techniques, widely adopted in edge ML, are directly applicable to FPGA-based quantum readout systems.

\subsection{Baseline CNN Model}

The original single-qubit CNN model \cite{Phuttitarn_2024} has approximately 355,000 trainable parameters. When scaled directly to multi-qubit arrays, the model complexity grows to over 74 million parameters, rendering real-time inference infeasible. 

\subsection{Pruned Model}

Our optimized single-qubit model eliminates unnecessary convolutional and feed-forward layers through strategic pruning, reducing the parameter count from 355k to approximately 5k---a 70$\times$ reduction---while preserving high readout fidelity.

Key insights driving this reduction include the removal of fully connected layers, which contributed minimally to classification accuracy but dominated the parameter count and are extremely expensive to implement on FPGAs. In addition, we pruned convolutional filter blocks, based on empirical analysis showing that only a small subset of filters contribute meaningfully to the model's output. This streamlined design not only reduces memory and compute demands but also simplifies deployment on FPGAs by minimizing latency-critical operations.

\subsection{Hierarchical Multi-Qubit Model}

To scale effectively, we introduce a hierarchical CNN architecture that reuses our lightweight single-qubit CNN as a modular component. Higher-level layers are added selectively to account for crosstalk between adjacent qubits. This hierarchical design enables high-fidelity multi-qubit readout with just ~18k parameters---a 4,000$\times$ reduction compared to the baseline CNN network. 

\begin{table}[htbp]
\caption{Accuracy and crosstalk fidelity comparison between baseline and optimized hierarchical CNN models. The hierarchical model achieves comparable accuracy with significantly improved nearest-neighbor crosstalk fidelity and 4000× fewer parameters.}
\centering
\begin{tabular}{l@{\hskip 6pt}c@{\hskip 6pt}c@{\hskip 6pt}c}
\hline \hline
\textbf{Model} & \textbf{C-NN} & \textbf{E-E} & \textbf{Mean Fidelity} \\
\hline
CNN-Site                        & 0.0134 & 0.0087 & 0.9856 \\
\textbf{Pruned CNN-Site}         & \textbf{0.0163} & \textbf{0.0061} & \textbf{0.9852} \\
CNN-Array                       & 0.0060 & 0.0058 & 0.9866 \\
\textbf{Pruned CNN-Array} & \textbf{0.0084} & \textbf{0.0077} & \textbf{0.9821} \\
\hline \hline
\end{tabular}
\label{tab:cnn_fidelity_comparison}
\end{table}

As shown in Table~\ref{tab:cnn_fidelity_comparison}, our pruned CNN-Array (18k model) achieves readout accuracy comparable to the baseline CNN single-site model (0.9821 vs. 0.9856) while significantly improving nearest-neighbor crosstalk fidelity (0.0084 vs. 0.0134). This demonstrates the benefit of explicitly modeling crosstalk at the architectural level while keeping the model size orders of magnitude smaller.

We outperform the baseline single-site CNN model at most readout durations, shown in Figure~\ref{fig:pruned_infidelity}a, while the hierarchical model occasionally outperforms the baseline multi-qubit CNN-array as shown in Figure~\ref{fig:pruned_infidelity}b. An overview of the hierarchical model design is illustrated in Figure~\ref{fig:pruned_infidelity}c.

While the reduced-complexity model improves upon the baseline CNN model, it is still substantially more complicated than the matched-filter model presented in the main text.

% \begin{sloppypar}
\bibliography{main}
% \end{sloppypar}

\end{document}